\documentclass{ws-ijmpe}

\begin{document}

\markboth{W. Trautmann et al.}{The Symmetry Energy in Nuclear Reactions}

\catchline{}{}{}{}{}

\title{THE SYMMETRY ENERGY IN NUCLEAR REACTIONS\\}

\author{\footnotesize W. TRAUTMANN\footnote{w.trautmann@gsi.de}, S. BIANCHIN, 
A.S.~BOTVINA\footnote{On leave from Inst. for Nuclear Research, Russian Academy of Sciences, 117312 Moscow, Russia}, A.~LE~F\`EVRE, Y.~LEIFELS, 
C.~SFIENTI\footnote{Present address: Universit\`{a} di Catania and INFN-Sezione di Catania, I-95123, Catania, Italy}, AND THE ALADIN COLLABORATION}
\address{GSI Helmholtzzentrum f\"{u}r Schwerionenforschung GmbH, \\
Planckstr. 1, D-64291 Darmstadt, Germany
}
\author{N. BUYUKCIZMECI, R. OGUL}
\address{Department of Physics, University of Sel\c{c}uk, 42079 Konya, Turkey} 
\author{I.N. MISHUSTIN}
\address{Frankfurt Institute for Advanced Studies, J.W. Goethe University,\\
D-60438 Frankfurt am Main, Germany}
\author{M. CHARTIER, P.Z. WU}
\address{University of Liverpool, Liverpool L69 7ZE United Kingdom}
\author{R.C. LEMMON}
\address{STFC Daresbury Laboratory, Warrington, WA4 4AD United Kingdom}
\author{Q. LI}
\address{School of Science, Huzhou Teachers College, Huzhou 313000, China}
\author{J. {\L}UKASIK, P. PAW{\L}OWSKI}
\address{H. Niewodnicza{\'n}ski Institute of Nuclear Physics, Pl-31342 Krak{\'o}w, Poland}
\author{A. PAGANO, P. RUSSOTTO}
\address{INFN-Sezione di Catania and LNS, I-95123 Catania, Italy}

\maketitle

\begin{history}
\received{(received date)}
\revised{(revised date)}
\end{history}

\begin{abstract}
New results for the strength of the symmetry energy are presented which illustrate
the complementary aspects encountered in reactions probing nuclear densities below 
and above saturation.
A systematic study of isotopic effects in 
spectator fragmentation was performed at the ALADIN spectrometer with 
$^{124}$Sn primary and $^{107}$Sn and $^{124}$La secondary beams of 600 MeV/nucleon 
incident energy. The analysis within the 
Statistical Fragmentation Model shows that the symmetry-term coefficient 
entering the liquid-drop description of the emerging fragments decreases 
significantly as the multiplicity of fragments and light particles from the
disintegration of the produced spectator systems increases. 
Higher densities were probed in the FOPI/LAND study of nucleon and light-particle
flows in central and mid-peripheral collisions 
of $^{197}$Au+$^{197}$Au nuclei at 400 MeV/nucleon incident energy.
From the comparison of the measured neutron and hydrogen squeeze-out ratios with
predictions of the UrQMD model a moderately soft symmetry term with a density dependence 
of the potential term proportional to ($\rho/\rho_0)^{\gamma}$ with $\gamma = 0.9 \pm 0.3$ 
is favored.
\end{abstract}

\section{Introduction}

Microscopic calculations are rather consistent in their predictions for the
strength of the symmetry energy at nuclear densities below saturation.\cite{baldo04}
For homogeneous matter, the symmetry energy decreases with decreasing density, roughly in
proportion to $(\rho/\rho_0)^{\gamma}$ where $\rho_0$ represents the nuclear density 
at saturation and $\gamma$ is approximately 2/3. The tendency of nuclear matter to cluster 
modifies this result, however. The symmetry energy is expected to be finite at 
very low densities, mainly because of $\alpha$-particle formation,\cite{horo06,kowa07}
and should reflect the abundance of larger fragments at the freeze-out conditions of
multifragmentation reactions. In the Statistical Multifragmentation Model 
which has been successfully applied to many types of fragmentation reactions, 
this is accounted for by using, in its standard version,
the symmetry-term coefficient of Weizs\"{a}cker's formula for stable nuclei 
for the description of the emerging fragments.\cite{bond95} 
Recent experiments have shown, however, that this scenario of an idealized freeze-out 
will have to be modified in order to be consistent with isotope distributions and other 
observables derived from isotopically resolved fragment yields (for references, 
see reviews\cite{dyntherm,lipr08} and contributions to this proceedings). 
The excited fragments seem to be modified in the hot environment which has important
consequences for astrophysical processes such as the collapse and explosion of massive
stars.\cite{botv04} 

At higher-than-normal densities, on the other hand, our knowledge regarding the
strength of the symmetry energy is much less complete.\cite{dyntherm,lipr08} 
Most of the available results are obtained by extrapolating from below or
near saturation to higher densities, thereby relying on a power law 
parameterization ($\rho/\rho_0)^{\gamma}$ for the density dependence of the potential 
part of the symmetry term. According to present findings, the density dependence is 
moderately soft, with $\gamma$ in the range of 0.5 to 1.0,
and appears to be rather consistent among different observables.\cite{klimk07,tsang09} 
Very recently, a super-soft symmetry term was deduced from the analysis\cite{xiao09} of
the charged-pion ratios for $^{197}$Au+$^{197}$Au reactions at 400 MeV/nucleon
incident energy measured by the FOPI collaboration.\cite{reis07} 
This result is of particular interest because of its 
consequences for neutron-star properties (see, e.g., Refs.\cite{kutsch94,wen09}) but
obviously contradicts the extrapolations. To clarify the situation, it will be
necessary to employ additional probes sensitive to the high-density stage of the 
reaction and capable of constraining the model analyses and the deduced
parameters of the equation of state of neutron-rich matter. 

The new results presented in the following, obtained from experiments conducted at the 
GSI laboratory with beams of several hundreds of MeV per nucleon, address these 
currently discussed aspects. From an isoscaling analysis of spectator fragmentation, a 
considerable modification of the symmetry term coefficient representing the
fragment properties in the hot freeze-out environment is deduced, and the ratio
of neutron versus hydrogen squeeze-outs is shown to be a sensitive observable for  
determining the symmetry energy at supra-saturation densities.

\section{Isoscaling in Spectator Fragmentation}

\begin{figure}[th]
\centerline{\psfig{file=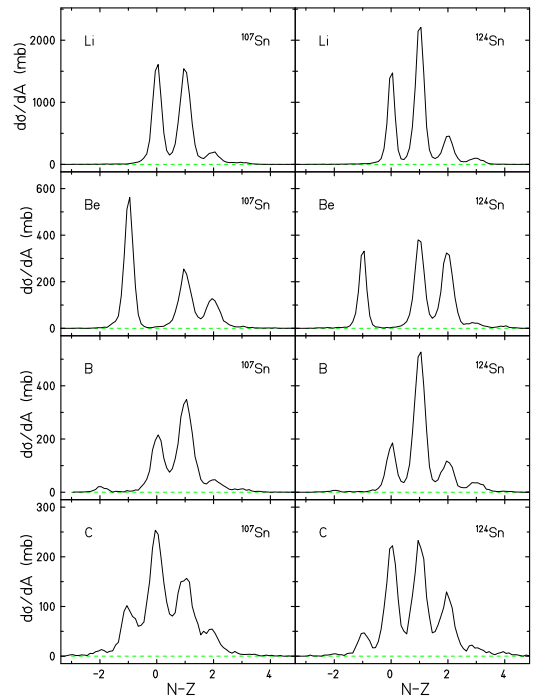,width=7cm}}
\vspace*{8pt}
\caption{Inclusive mass distributions for fragments with $3 \le Z \le6$ obtained for
$^{107}$Sn (left panels) and $^{124}$Sn projectiles (right panels).}
\label{fig:mass}
\end{figure}

The ALADIN experiment S254, conducted in 2003 at the SIS heavy-ion synchrotron,
was aimed at studying isotopic effects in projectile 
fragmentation at relativistic energies.\cite{traut08,sfienti09} 
Besides stable $^{124}$Sn beams,
neutron-poor secondary $^{107}$Sn and $^{124}$La beams were used in order to 
explore a wide range of isotopic compositions. 
The radioactive beams were produced  at the fragment 
separator FRS\cite{frs92} by the fragmentation of primary $^{142}$Nd 
projectiles with energies near 900 MeV/nucleon in a thick beryllium  target. 
The FRS was set to select $^{124}$La and, in the second part of the experiment, 
$^{107}$Sn projectiles which were then delivered to the experiment. 
All three beams had a laboratory energy of 600 MeV/nucleon and were directed
onto $^{\rm nat}$Sn targets with an areal density of 500 mg/cm$^2$.
The acceptance for fragments from the projectile-spectator decays was 90\% 
for fragments with atomic number $Z=3$ and increased gradually to about 95\% for 
$Z \ge 6$. The obtained mass resolution was 7\% (FWHM) for $Z \le 3$, 
decreasing to 3\% for $Z\geq 6$, so that masses for $Z \leq 10$ were individually 
resolved (Fig.~\ref{fig:mass}).

\begin{figure}[th]
\centerline{\psfig{file=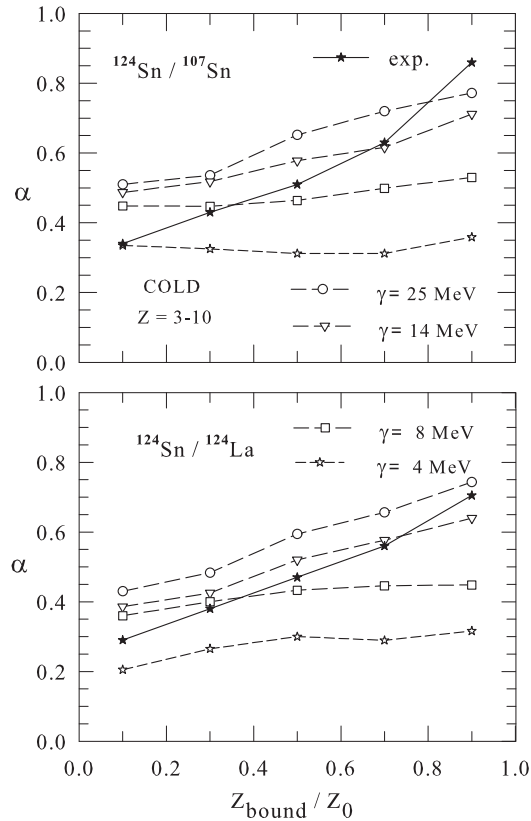,width=7cm}}
\vspace*{8pt}
\caption{Experimental data (stars) and SMM ensemble calculations (open symbols) of
isoscaling coefficients $\alpha$ extracted from yield ratios of fragments 
($3 \le Z \le 10$) from 
$^{124}$Sn and $^{107}$Sn projectiles (top panel), and from $^{124}$Sn and $^{124}$La
projectiles (bottom), versus $Z_{\rm bound}$ normalized with respect to the nominal
projectile charges $Z_0 = 50$ and 57.
Four symmetry-term coefficients~$\gamma$ were used in the SMM calculations as indicated
in the figure.}
\label{fig:alpha}
\end{figure}

In order to reach the necessary beam intensity of about 1000 particles/s with
the smallest possible mass-to-charge ratio $A/Z$, it was found necessary to
accept a distribution of neighbouring nuclides together with the requested
$^{124}$La or $^{107}$Sn isotopes. 
Their mean compositions were $<$$Z$$>$ = 56.8 (49.7) 
and $<$$A/Z$$>$ = 2.19 (2.16) for the nominal $^{124}$La ($^{107}$Sn) beams, 
respectively.\cite{luk08} Model studies consistently predict that these 
$<$$A/Z$$>$ values are also representative for the spectator systems emerging 
after the initial cascade stage of the reaction. In particular, the differences 
in $<$$A/Z$$>$ between the neutron-rich and neutron-poor cases are expected to remain the same 
within a few percent.\cite{botv02,lef05}

Global fragmentation observables were found to depend only weakly on the isotopic 
composition.\cite{traut08,sfienti09} 
This includes, in particular, the mean multiplicity of intermediate-mass fragments 
($3 \le Z \le 20$), the largest atomic number $Z_{\rm max}$ within a partition, and
the evolution of these quantities with $Z_{\rm bound}$. The sorting variable 
$Z_{\rm bound} = \Sigma Z_i$ of fragments with $Z_i \ge 2$ represents the atomic number 
$Z$ of the spectator system, apart from emitted hydrogen isotopes, and is inversely
correlated with the transferred excitation energy.

\begin{figure}[th]
\centerline{\psfig{file=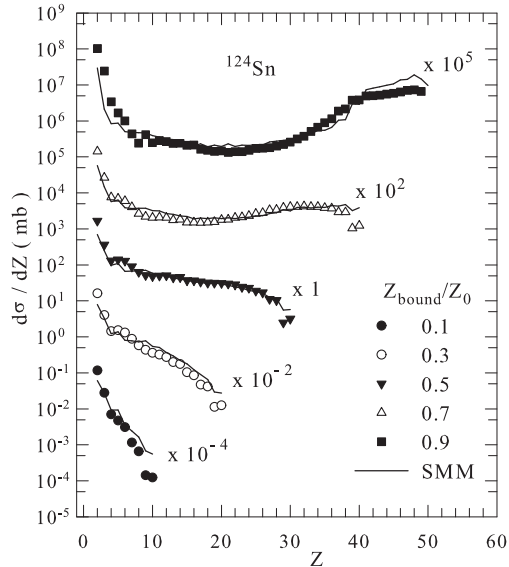,width=7cm}}
\vspace*{8pt}
\caption{Measured cross sections for the fragment production in selected $Z_{\rm bound}$ 
intervals of 10-unit width as a function of the fragment $Z$ (symbols) in comparison with 
the results of SMM ensemble calculations (lines) for the case of $^{124}$Sn projectiles. 
The centers of the five $Z_{\rm bound}$ intervals and the scaling factors used for displaying
the cross sections are indicated in the figure.}
\label{fig:charge}
\end{figure}

The comparison of the measured fragment yields from neutron-rich and neutron-poor
systems shows that isoscaling is observed.\cite{bianchin_bormio} The isoscaling
parameter $\alpha$, determined from the yields for $3 \le Z \le 10$ 
is found to decrease rapidly as the disintegration of the
spectator systems into fragments and light particles increases (Fig.~\ref{fig:alpha}), 
confirming earlier results for the fragmentation of target spectators in reactions of
$^{12}$C on $^{112,124}$Sn at 300 and 600 MeV/nucleon.\cite{lef05} Nearly identical 
results are obtained for the isotopic and isobaric pairs of reactions.

The analysis of the data with the Statistical Multifragmentation
Model\cite{bond95} was performed with ensemble calculations adapted to the
participant-spectator scenario at relativistic energies.\cite{botv95} The ensembles 
of excited systems with varying excitation energy and mass were chosen so as to best
reproduce the charge spectra and correlations observed for the fragment
production (Fig.~\ref{fig:charge}). 
The mean neutron-to-proton ratios $<$$N$$>$/$Z$ for intermediate-mass fragments up to
$Z=10$ and the isoscaling parameters were found to be particularly sensitive to the 
coefficient $\gamma$ of the symmetry term $E_{\rm sym}(A,Z)=\gamma (A-2Z)^2/A$ in 
the liquid-drop description of excited fragments at 
freeze-out.\cite{ogul09,gamma} For the isoscaling parameter $\alpha$, this is demonstrated in 
Fig.~\ref{fig:alpha}. The SMM standard value $\gamma = 25$~MeV is applicable only in the bin
of largest $Z_{\rm bound}$. Smaller values have to be chosen for reproducing the rapidly
decreasing parameter $\alpha$ in the fragmentation regime at smaller $Z_{\rm bound}$.

\begin{figure}[th]
\centerline{\psfig{file=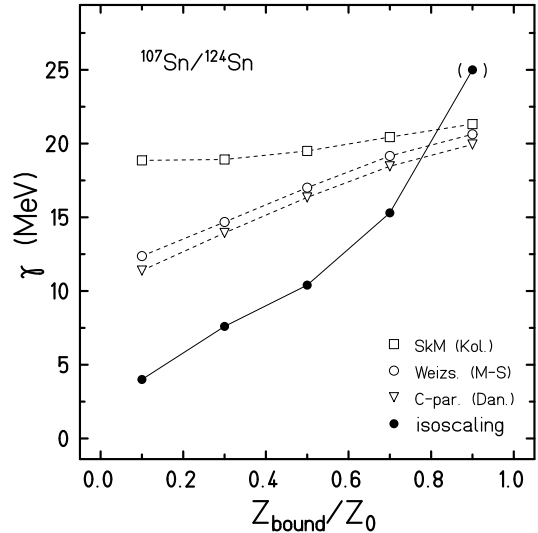,width=7cm}}
\vspace*{8pt}
\caption{Symmetry term coefficient $\gamma$ as expected from the changing
fragment-mass distributions using surface and volume symmetry-term coefficients
from Refs.\protect\cite{ms,dani09,kol08} (open symbols) in comparison with 
the coefficient $\gamma$ for hot fragments obtained from the isoscaling
analysis for the $^{107,124}$Sn pair of reactions 
with the Statistical Multifragmentation Model (dots).}
\label{fig:gamma}
\end{figure}

The symmetry-term coefficient $\gamma$ for hot fragments resulting from the 
isoscaling analysis for the $^{107,124}$Sn pair of reactions is
shown in Fig.~\ref{fig:gamma} as a function of $Z_{\rm bound}/Z_0$. For comparison,
three predictions are shown in the figure, obtained with rather different approaches but
all containing the effect of the surface-symmetry term whose importance increases for the 
lower-mass fragments. The coefficients of the mass formula of Myers and Swiatecki are
adapted to ground-state masses.\cite{ms} From the energies of isobaric analog 
states, a relation between the volume and surface capacitances of nuclei for absorbing
asymmetry $N-Z$ was derived by Danielewicz and Lee,\cite{dani09} while Kolomietz and Sanzhur
have used a variational approach using Skyrme forces to derive equilibrium values for the
volume symmetry term with surface and curvature corrections for nuclei along the 
$\beta$-stability line.\cite{kol08} With these coefficients, the effective symmetry energy 
averaged over the set of partitions was calculated for the five bins in $Z_{\rm bound}$
after the experimental $Z$ distributions had been converted to mass distributions using 
the projectile $N/Z$. The obtained results show similar trends. 
The smaller fragments produced at higher excitations cause the effective mean 
symmetry term to decrease with decreasing $Z_{\rm bound}$ in all three cases 
but at a slower rate than that resulting from the isoscaling analysis of the 
experimental yield ratios.

\section{Neutron and Hydrogen Flows}

\begin{figure}[th]
\centerline{\psfig{file=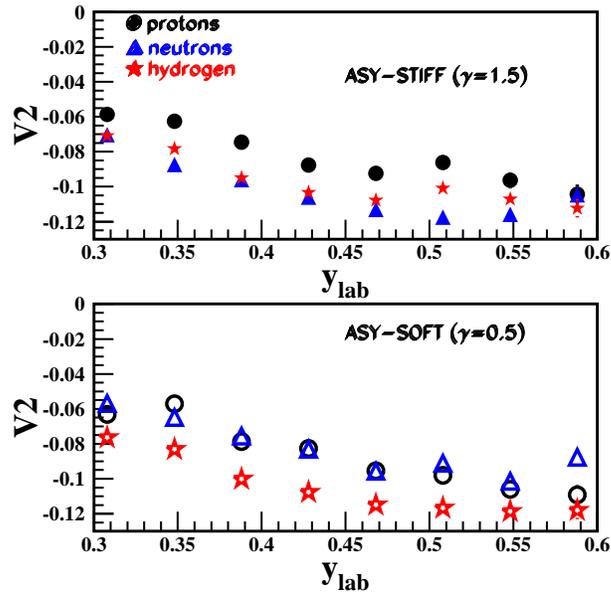,width=8cm}}
\vspace*{8pt}
\caption{Elliptic flow parameter $v_2$ for mid-peripheral $^{197}$Au+$^{197}$Au 
collisions at 400 MeV per nucleon as calculated with the UrQMD model for protons (circles), neutrons (triangles), and the total hydrogen yield (stars) as a function of the laboratory rapidity $y_{\rm lab}$. The results have been filtered to correspond to the geometrical acceptance of the LAND setup used in the joined experiment. The predictions obtained with a stiff and a soft density dependence of the symmetry term are given in the upper and lower panels, respectively.
}
\label{fig:v2cut}
\end{figure}

In two experiments at GSI combining the LAND and FOPI (Phase 1) detectors, 
neutron and 
hydrogen collective flow observables from $^{197}$Au+$^{197}$Au collisions at 
400, 600, and 800 MeV/nucleon have been measured.\cite{leif93,lamb94} This data set is 
presently being reanalyzed in order to determine optimum conditions 
for a dedicated new experiment,\cite{s394} but also with the aim to produce constraints
for the symmetry energy at the high densities probed in central collisions at these energies.
The results reported here are obtained by comparing with predictions
of the UrQMD model which has recently been adapted to heavy ion reactions 
at intermediate energies.\cite{qli05}

The predictions obtained for the elliptic flow of neutrons, protons, and 
hydrogen yields for $^{197}$Au+$^{197}$Au at 400 A MeV are shown in Fig. \ref{fig:v2cut}.
Two values are chosen for the power-law exponent describing the density dependence 
of the potential part of the symmetry energy, $\gamma = 1.5$ (asy-stiff) and 
$\gamma = 0.5$ (asy-soft). The UrQMD outputs have been filtered in order to
correspond to the geometrical acceptance 
of the FOPI/LAND experiment. This produces the asymmetry of $v_2$ with respect to
mid-rapidity $y_{\rm lab}=0.448$ because higher transverse momenta are selected with
increasing rapidity. 
The neutron squeeze-out is significantly larger (larger absolute value of $v_2$)
in the asy-stiff case 
(upper panel) than in the asy-soft case (lower panel) while the proton and hydrogen 
flows respond only weakly to the variation of $\gamma$ within the chosen interval. 

\begin{figure}[th]
\centerline{\psfig{file=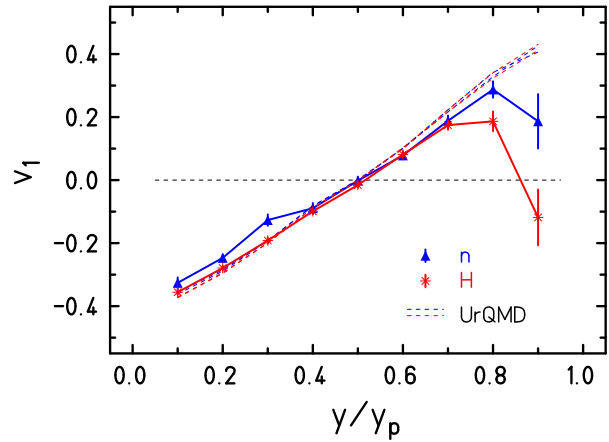,width=8cm}}
\vspace*{8pt}
\caption{Directed-flow parameters $v_1$ for neutrons (triangles) and 
hydrogen isotopes (stars) for mid-peripheral
($5.5<b<7.5$ fm) collisions of $^{197}$Au+$^{197}$Au at 400 MeV/nucleon as a 
function of the scaled rapidity $y/y_p$ in comparison with 
the UrQMD predictions for $\gamma = 1.5$ (a-stiff) and $\gamma = 0.5$ (a-soft)
represented by the dashed lines.}
\label{fig:v1}
\end{figure}

A nearly negligible sensitivity to the stiffness of the symmetry energy 
is exhibited by the directed flow, according to the UrQMD model. In Fig.~\ref{fig:v1}, 
the first Fourier coefficient $v_1$ of the azimuthal particle distribution 
with respect to the reconstructed reaction plane is shown as a function
of the rapidity $y$, normalized with respect to the projectile rapidity $y_0 = 0.896$, and 
for the interval of impact parameters $5.5$~fm~$<b<7.5$~fm. The
dashed lines representing the results for neutrons and for hydrogens and for asy-soft and
asy-stiff parameterizations of the symmetry energy fall practically on top of each other. 
The predictions compare well, however, with the experimental results for the multiplicity 
bin PM3 (see Ref.\cite{leif93}) expected to correspond to this range of mid-peripheral 
impact parameters (Fig.~\ref{fig:v1}). The range of rapidities
$y \ge 0.8$ at which the data deviate from the expected linearity coincides with the
shift of the LAND acceptance to transverse momenta $p_t \ge 0.75$~GeV/c/nucleon at which
the yields start to drop.

\begin{figure}[th]
\centerline{\psfig{file=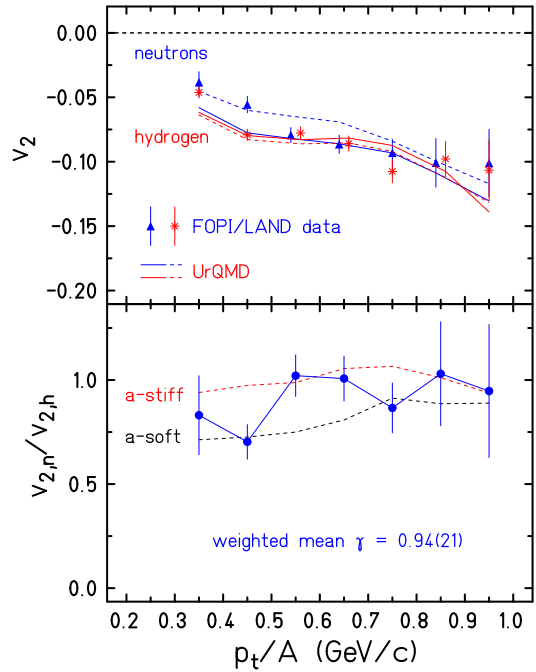,width=7cm}}
\vspace*{8pt}
\caption{Elliptic flow parameters $v_2$ for neutrons (triangles) and 
hydrogen isotopes (stars, top panel) and their ratio $v_{2,n}/v_{2,h}$ 
(bottom panel, dots) for 
central ($b<7.5$ fm) collisions of $^{197}$Au+$^{197}$Au at 400 MeV/nucleon as a 
function of the transverse momentum per nucleon $p_t/A$ in comparison with 
the UrQMD predictions for $\gamma = 1.5$ (a-stiff) and $\gamma = 0.5$ (a-soft)
represented by the dashed lines.}
\label{fig:ptdep}
\end{figure}

The results obtained for the second Fourier coefficient $v_2$ describing elliptic flow
are shown in Fig.~\ref{fig:ptdep}. Here the combined data set for central and mid-peripheral 
collisions (PM3 to PM5) is used and compared to the corresponding UrQMD predictions for 
$b<7.5$~fm. The global rise of the absolute magnitude of $v_2$ with $p_t$ is well described 
even though the approximately 15\% correction for the
dispersion of the reaction plane (cf. Ref.\cite{andro06}) is not applied to the data. 
In contrast to the directed flow and as expected from Fig.~\ref{fig:v2cut}, 
the squeeze-out of neutrons depends significantly on the symmetry term chosen for the 
calculations. It is considerably weaker in the asy-soft case while the predictions
for the asy-stiff case and for the hydrogen isotopes practically 
coincide (Fig.~\ref{fig:ptdep}, upper panel).

For a quantitative evaluation, the ratio of neutron-over-hydrogen flows is
proposed as an observable which should be insensitive to
uncertainties resulting from the experimental determination of the reaction
plane and from the matching of the experimental and theoretical impact-parameter
intervals. The comparison shows that the experimental ratios scatter in-between
the predictions for the asy-soft and asy-stiff cases (Fig.~\ref{fig:ptdep}, lower panel).
A linear interpolation between these predictions, averaged over 
$0.3 < p_t/A \le 1.0$ GeV/c, yields $\gamma = 0.94 \pm 0.21$.
A smaller but within errors consistent value $\gamma = 0.52 \pm 0.30$ is obtained
if the comparison is restricted to mid-peripheral impact-parameters 
$5.5 \le b<7.5$~fm.\cite{traut09}
The power law coefficients depend weakly on the symmetry energy at saturation chosen 
in the parameterization. The slight increase of $\gamma$ if this value is lowered 
confirms that densities above saturation are probed.
Other systematic uncertainties have been found to remain within 
$\Delta\gamma \approx 0.2$. Together with the kinetic term proportional to 
$(\rho/\rho_0)^{2/3}$, the squeeze-out data thus indicate a moderately soft behavior of the 
symmetry energy at supra-saturation densities. 

This result can be considered as, within errors, consistent with the density dependence 
deduced from fragmentation experiments probing nuclear matter near or below 
saturation\cite{tsang09} and with the slightly softer density dependence resulting from 
the analysis of the pygmy dipole resonance in heavy nuclei.\cite{klimk07} 
It also shows that the super-soft density dependence resulting from the IBUU 
analysis\cite{xiao09} of the $\pi ^-/\pi ^+$ yield ratios can, at this time, not be
considered as a unique consequence of experiments probing higher densities. 
Pion yields are expected to be produced during the high-density phase of the 
reaction\cite{zhang09} but are also subject to considerable
in-medium effects which have to be controlled with high precision.\cite{ko09} 
More work, possibly including additional observables and focussing on the consistency 
among them, will be needed in order to arrive at firm conclusions regarding the symmetry 
energy at high density. In view of the far-reaching and mani-fold consequences, this
task deserves highest priority.

Illuminating discussions with Lie-Wen Chen, Bao-An Li, M.~Di Toro, and H.H.~Wolter are 
gratefully acknowledged. This work has been supported by the European Community under 
contract No. HPRI-CT-1999-00001 and FP7-227431 (HadronPhysics2) .


\begin{thebibliography}{99}
\bibitem{gamma}
note that the same notation $\gamma$ has been adopted for the symmetry term coefficient 
of excited fragments in the SMM and for the power-law coefficient
describing the density dependence of the symmetry energy.
\end{thebibliography}
\end{document}